\documentclass[aps,prl,english,twocolumn,showpacs]{revtex4}
\usepackage{float}
\usepackage{graphicx}

\makeatletter

\usepackage{babel}
\makeatother
\begin{document}
\def\eqd{\,{\buildrel d \over =}\,}\def\subst{\,{\buildrel {y'=y f(\tau^{-1})} \over =}\,}\textsf{}

\title{Self-similar signature of the active solar corona within the inertial range
of solar wind turbulence}

\author{K. Kiyani}

\email{k.kiyani@warwick.ac.uk}

\author{S. C. Chapman}

\author{B. Hnat}

\author{R. M. Nicol}

\affiliation{Centre for Fusion, Space and Astrophysics; Dept. of Physics, University
of Warwick, Coventry CV4 7AL, UK}

\date{\today}

\begin{abstract}

We quantify the scaling of magnetic energy density in the inertial range of solar wind turbulence seen in-situ at 1AU with respect to solar activity. At solar maximum, when the coronal magnetic field is dynamic and topologically complex, we find self-similar scaling in the solar wind, whereas at solar mimimum, when the coronal fields are more ordered, we find multifractality. This quantifies the solar wind signature that is of direct coronal origin, and distinguishes it from that of local MHD turbulence, with quantitative implications for coronal heating of the solar wind.
\end{abstract}


\maketitle The interplanetary solar wind exhibits fluctuations
characteristic of Magnetohydrodynamic (MHD) turbulence evolving in
the presence of structures of coronal origin. In-situ spacecraft
observations of plasma parameters are at minute (or below)
resolution for intervals spanning the solar cycle, and provide a
large number of samples for statistical studies. These reveal a
magnetic Reynolds number  $\sim10^{5}$ \cite{Matthaeus2005} and
power spectra with a clear inertial range over several orders of
magnitude characterised by a power law Kolmogorov exponent of $\sim -5/3$.
Quantifying the properties of fluctuations in
the solar wind can thus provide insights into MHD turbulence 
and also inform our understanding of coronal processes
and ultimately, the mechanisms by which the solar wind is heated.
Quantifying these fluctuations is also central to understanding
the propagation of cosmic rays in the heliosphere \cite{Giacalone1999}.

Coronal heating mechanisms are studied in terms of the scaling
properties of coronal structures \cite{Schrijver1998,Tu2005},
heating rates \cite{Klimchuk1995} and diffusion via
random walks of magnetic field lines \cite{Giacalone2006}, all of
which suggest self-similar processes. The solar wind is also
studied in-situ to infer information pertaining to coronal
processes. Large scale coherent structures of solar origin, such as CMEs, can be
directly identified in the solar wind. At frequencies below the
`Kolmogorov- like' inertial range, the solar wind exhibits an
energy containing range which shows $\sim 1/f$ scaling \cite{Goldstein1984} \cite{Matthaeus1986}. 
Solar flares show scale invariance in their energy release statistics 
over several orders of magnitude \cite{Aschwanden2000} which 
has been discussed in terms of Self-Organized Criticality 
(SOC) \cite{Lu1991,Hughes2003}.
Within the inertial range, the observed solar wind magnetic
fluctuations are principally Alfv\'{e}nic in character with
asymmetric propagation anti-sunwards \cite{Horbury2005}. In-situ
plasma parameters which directly relate to cascade theories of
ideal incompressible MHD turbulence, such as velocity, magnetic
field, and the Elsasser variables have thus been extensively
studied in the solar wind (\cite{TuMarsch1995} and refs. therein). 
These show multifractal scaling in their higher order moments consistent with intermittent
turbulence \cite{Horbury1997,Pagel2001}. 
Intriguingly, the magnetic energy density $B^{2}$ and
number density $\rho$ show approximately \emph{self-similar}
scaling in the inertial range \cite{Hnat2002,Hnat2004}. These parameters
 are insensitive to Alfv\'{e}nicity, and  do not relate
directly to MHD cascade theories.

In this Letter we quantify the scaling seen in $B^2$ in the
inertial range of solar wind turbulence with respect to coronal
structure and dynamics. We employ a recently developed
technique \cite{Kiyani2006} that sensitively distinguishes between
self-similarity and multifractality in timeseries. This will allow
us to distinguish and quantify the solar wind
signature that is of direct coronal origin from that of local MHD
turbulence, with quantitative implications for our understanding
of coronal heating of the solar wind.

The WIND and ACE spacecraft spend extended intervals at $\sim$1 AU
in the ecliptic and provide in-situ magnetic field observations
of the solar wind over extended periods covering all phases of the
solar cycle. We focus on a comparison between
solar maximum when the coronal structure is highly variable with
topologically complex magnetic structure, with that at solar
minimum when the coronal magnetic structure is highly ordered. The most
magnetically ordered region of the corona is at the poles at solar
minimum and observations of the corresponding quiet, fast solar
wind are provided by the ULYSSES spacecraft. The four data sets,%
\footnote{obtained from http://cdaweb.gsfc.nasa.gov/ and 
http://ulysses-ops.jpl.esa.int/ulysses/archive/vhm\_fgm.html%
},
that we consider here are then \emph{a}.)  WIND 60 seconds
averaged MFI data at the solar maximum year of 2000 and \emph{b}.)
at the solar minimum year of 1996; \emph{c}.) ACE  64 seconds
averaged MFI data for the year 2000; and \emph{d}.) ULYSSES 60
seconds averaged VHM/FGM data for July and August 1995. Data sets
\emph{a}--\emph{c} consist of $\sim4.5\times10^{5}$ points; and
\emph{d} consists of $\sim8.5\times10^{4}$ data points. 
Intervals corresponding to magnetospheric bow shock crossings
for WIND were removed by comparison with %
\footnote{http://lepmfi.gsfc.nasa.gov/mfi/bow\_shock.html%
}. The ACE spacecraft orbits around the Earth-Sun L1 point and the
ULYSSES data was obtained for the North polar pass of 1995. All of
the above intervals  show a $\sim-5/3$ power law scaling
inertial range in the power spectra of $|B|$ over several decades
which is indicative of a well developed turbulent fluid.

\begin{figure}[H]
a.) \includegraphics[width=0.8\columnwidth,keepaspectratio]{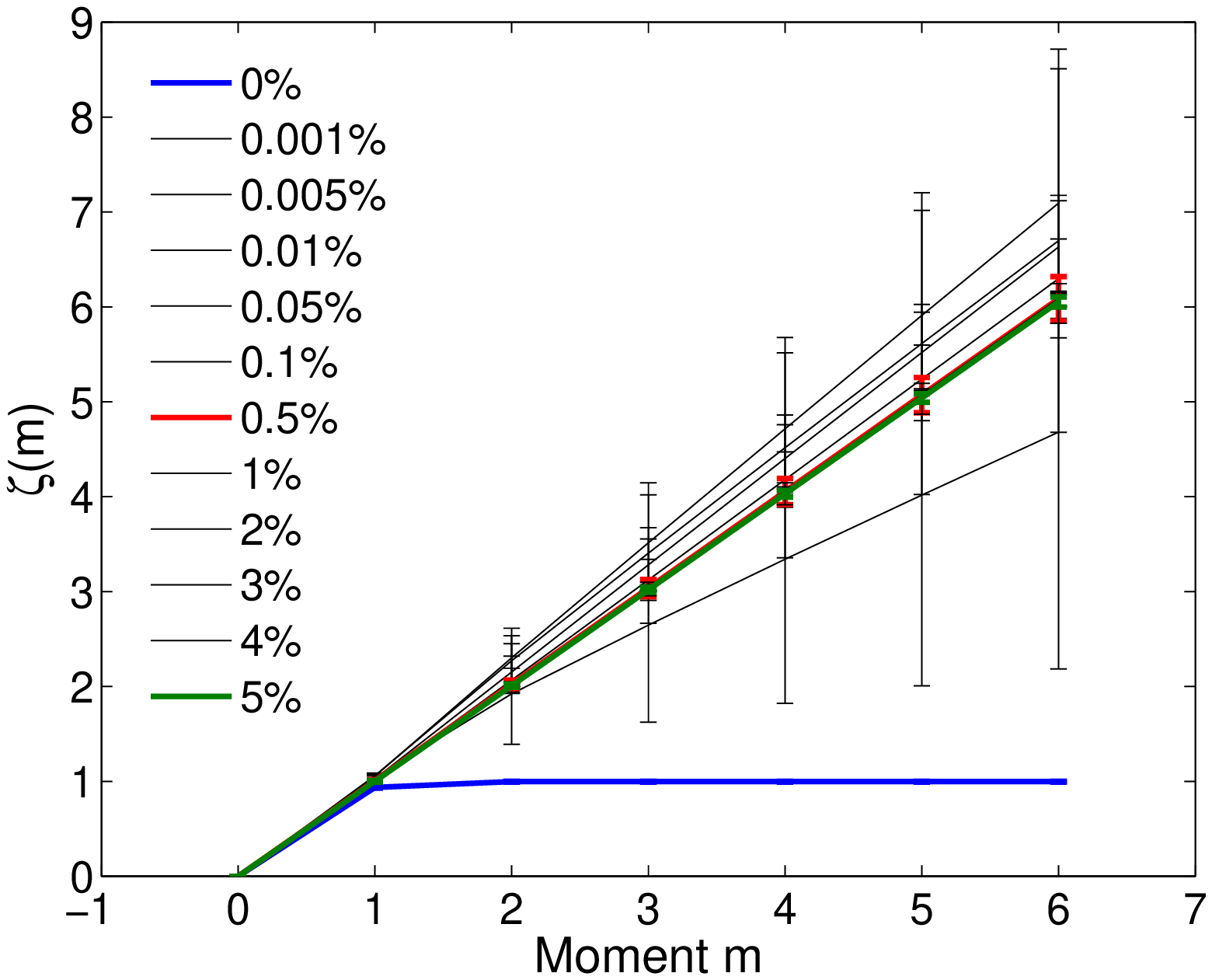}\newline
b.) \includegraphics[width=0.8\columnwidth,keepaspectratio]{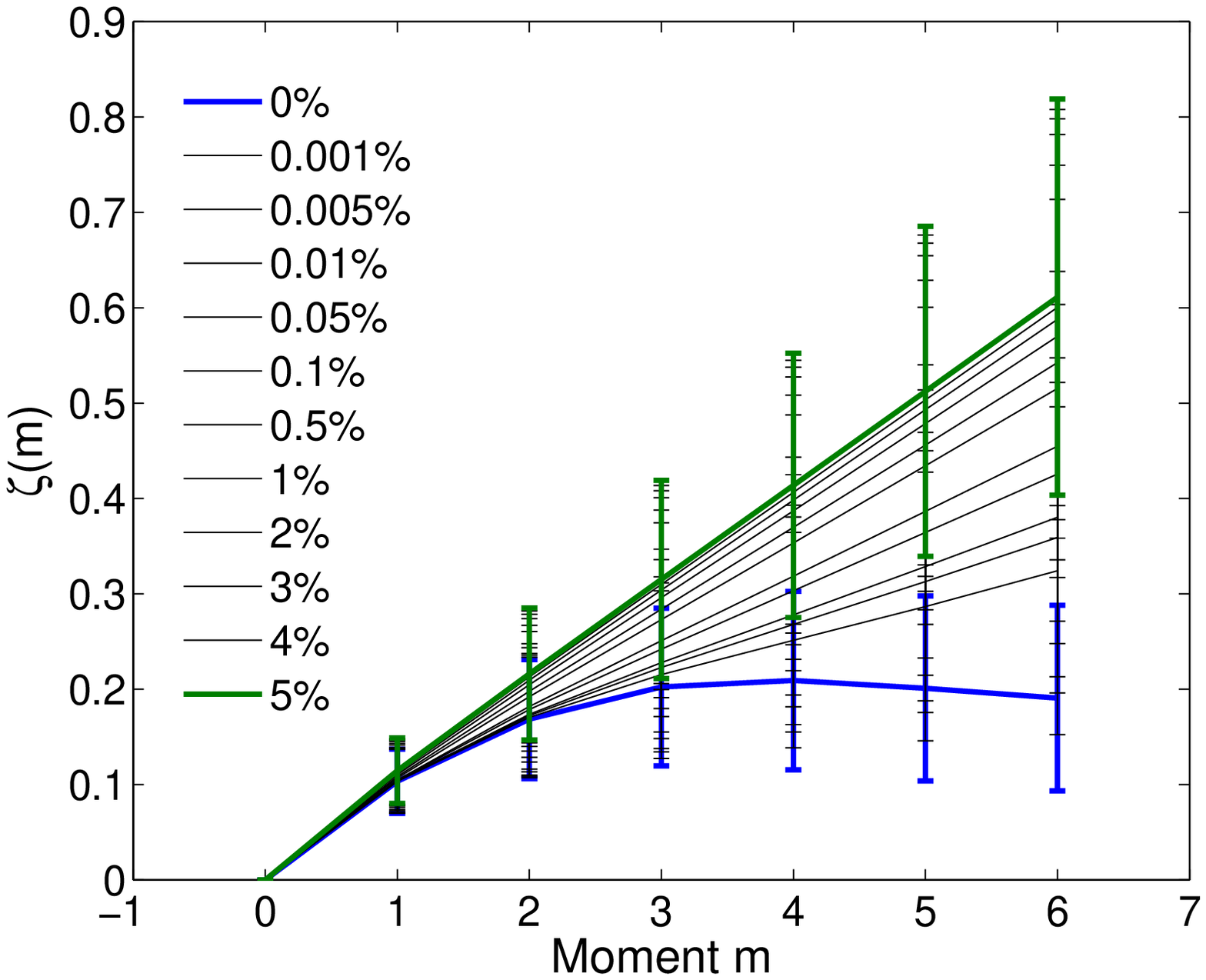}
\caption{\label{fig:zetapVsp}$\zeta(m)$ Vs. $m$ plots for a.) $\alpha=1.0$
symmetric L\'{e}vy process and b.) $p=0.6$ $p$-model process.}
\end{figure}

\begin{figure}
a.) \includegraphics[width=0.8\columnwidth,keepaspectratio]{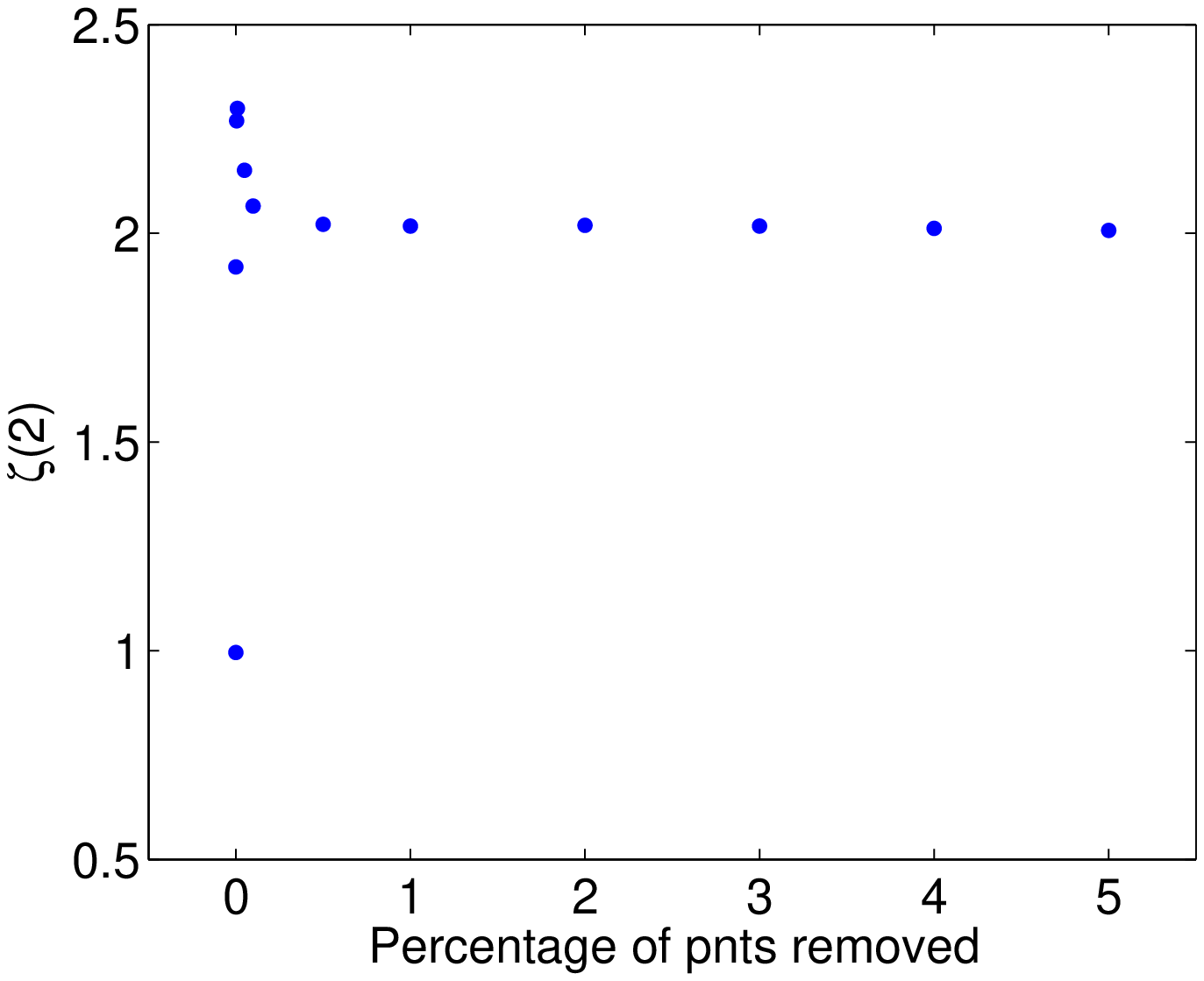}

b.) \includegraphics[width=0.8\columnwidth,keepaspectratio]{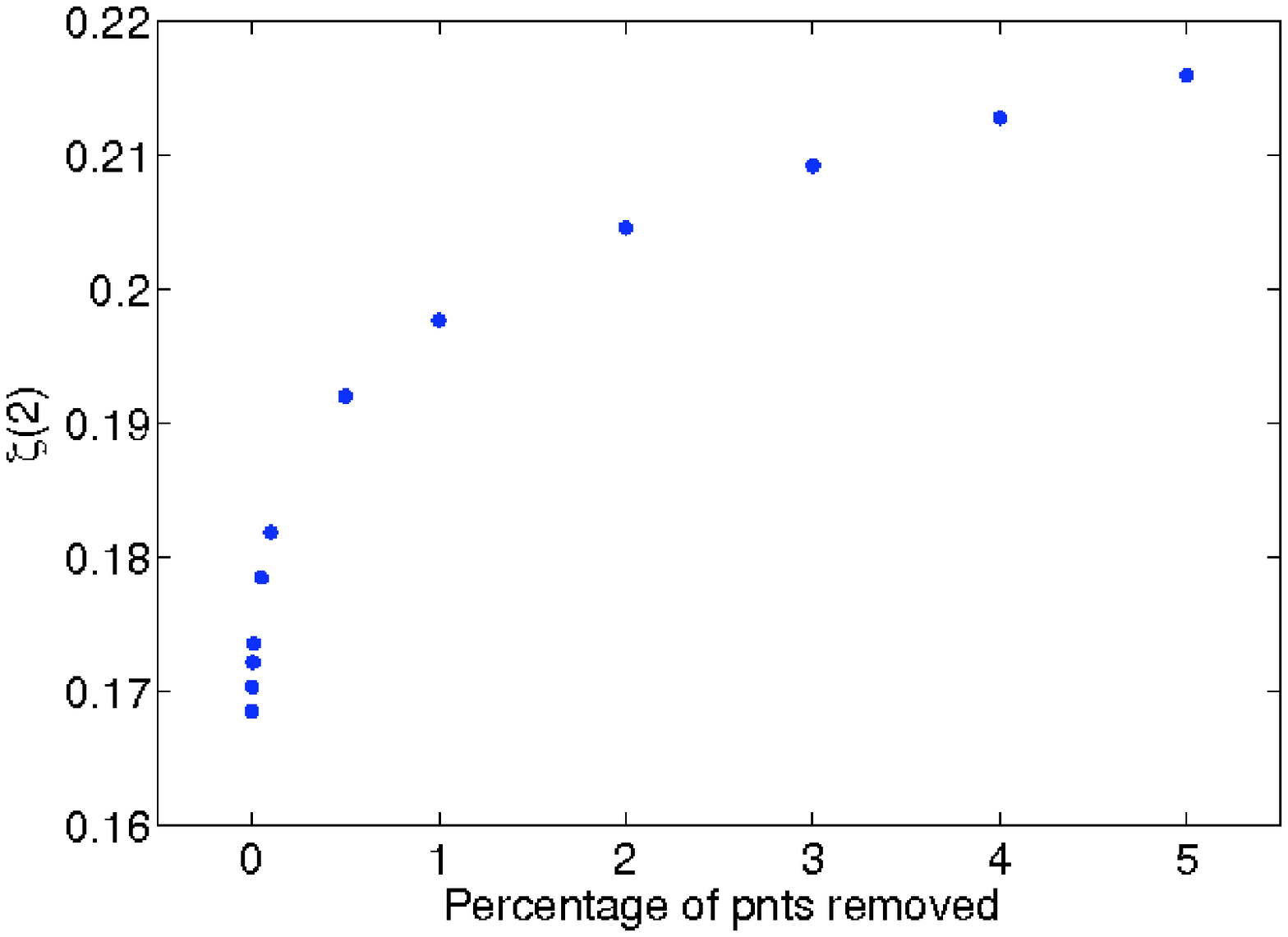}
\caption{\label{fig:zeta2Vspercsynthetic}Exponent of the second moment $\zeta(2)$
Vs. the percentage of points excluded for a.) the L\'{e}vy model
and b.) $p$-model.}
\end{figure}

We can access the statistical scaling properties of a timeseries
by constructing  differences
$y(t,\tau)=|B(t+\tau)|^{2}-|B(t)|^{2}$ on all available time
intervals $\tau$. The statistical scaling with
$\tau$ can be seen in the structure functions of order $m$ which
follows that of the moments of the PDF of $y$, $P(y,\tau)$:
\begin{equation} 
S^{m}(\tau)=\left\langle
\left|y\right|^{m}\right\rangle \
=\int_{-\infty}^{\infty}\left|y\right|^{m}P(y,\tau)dy\
,\label{eq:1}
\end{equation} 
where $\left\langle \right\rangle $
indicate ensemble averaging over $t$. Statistical self-similarity
 implies that any PDF at scale $\tau$ can be collapsed onto a
unique single variable PDF $\mathcal{P}_{s}$:
\begin{equation}
P(y,\tau)=\tau^{-H}\mathcal{P}_{s}(\tau^{-H}y)\
,\label{eq:2}
\end{equation}
where $H$ is the Hurst exponent.
Equation (\ref{eq:2}) implies that the increments $y$ are
\emph{self-affine} i.e. they obey the  statistical scaling
equality
$y(b\tau)\eqd b^{H}y(\tau)$
, such that the structure functions will
scale with $\tau$ as
\begin{equation}
S^{m}(\tau)=\tau^{\zeta(m)}\mathcal{S}_{s}^{m}(1)\
.\label{eq:3}
\end{equation} 
For the special case of a
statistically self-similar (random fractal) process,
$\zeta(m)=Hm$. This scaling with $H=1/3$ is characteristic of
Kolmogorov's 1941 theory of turbulence \cite{kolmogorov41a}, and  intermittency
corrections to this are modelled by quadratic and concave
$\zeta(m)$ (multifractals) \cite{Frisch1995}. A difficulty that can
arise in the experimental determination of the $\zeta(m)$ is that
for a finite length timeseries, the integral (\ref{eq:1}) is
not sampled over the range $(-\infty,+\infty)$; the outlying
measured values of $y$ determine the limits. In the case of a
heavy-tailed PDF the higher order moments (larger $m$) can yield a
$\zeta(m)$ that deviates strongly from the scaling of $P(y,\tau)$
in (\ref{eq:2}) \cite{Kiyani2006} (hereafter KCH). An operational
solution to this was demonstrated in KCH for a self-similar
process. Essentially one systematically excludes a minimal
percentage of the outlying events $y$ from the integral in
(\ref{eq:1}) so that the statistics of the PDF tails become well
sampled and the integral (\ref{eq:1}) yields a $\tau$ dependence
with the correct scaling of the self-similar process (\ref{eq:2}).
This method is sensitive in distinguishing self-affine scaling from weak multifractality.
 We illustrate this with two reference
models: the first of which is manifestly self-similar, an
$\alpha$-stable L\'{e}vy process of index $\alpha=1.0$ ($H=1/\alpha$)
\cite{Kiyani2006}; and the second, manifestly multifractal, i.e. a \emph{p}-model \cite{Meneveau1987} with $p=0.6$.
These synthetic data sets each consist of $10^{6}$ data points.
Figure \ref{fig:zetapVsp} shows plots of the exponents $\zeta(m)$
Vs. $m$ obtained from (\ref{eq:3}) by computing the gradients
of $\log S^{m}(\tau)$ for (a) the L\'{e}vy process and (b) the
multifractal model respectively. The exponents $\zeta(p)$ have
been recomputed as outlying data points are successively removed,
and we can see that removing a small fraction, $\sim 0.001\%$ of
the data leads to a large change in the computed $\zeta(p)$. A
reliable estimate of the exponents from the data requires rapid
convergence to robust values; shown in KCH to be a
property of self-affine timeseries. We can see this behaviour in
the L\'{e}vy model which quickly converges to linear
dependence of $\zeta(p)$ with $p$ as expected.
The multifractal $p$-model only begins to approach linearity after
$\sim3\%$ of the data is excluded. This apparent
linearity in the $p$-model is actually a divergence in the values
of the $\zeta(p)$. We see this behaviour if we plot
the value of  one of the exponents from Figure \ref{fig:zetapVsp}
versus the percentage of points removed. This is shown for
$\zeta(2)$ for the L\'{e}vy process (upper panel) and the
\emph{p}-model (lower panel) in Figure \ref{fig:zeta2Vspercsynthetic}.
As we successively exclude outlying data points, the
self-affine L\'{e}vy process quickly reaches a constant value for
$\zeta(2)=2/\alpha=2.0$ whereas for the multifractal, the
$\zeta(2)$ exponent shows a continuing secular drift.
\begin{figure}
a.)\includegraphics[width=0.8\columnwidth,keepaspectratio]{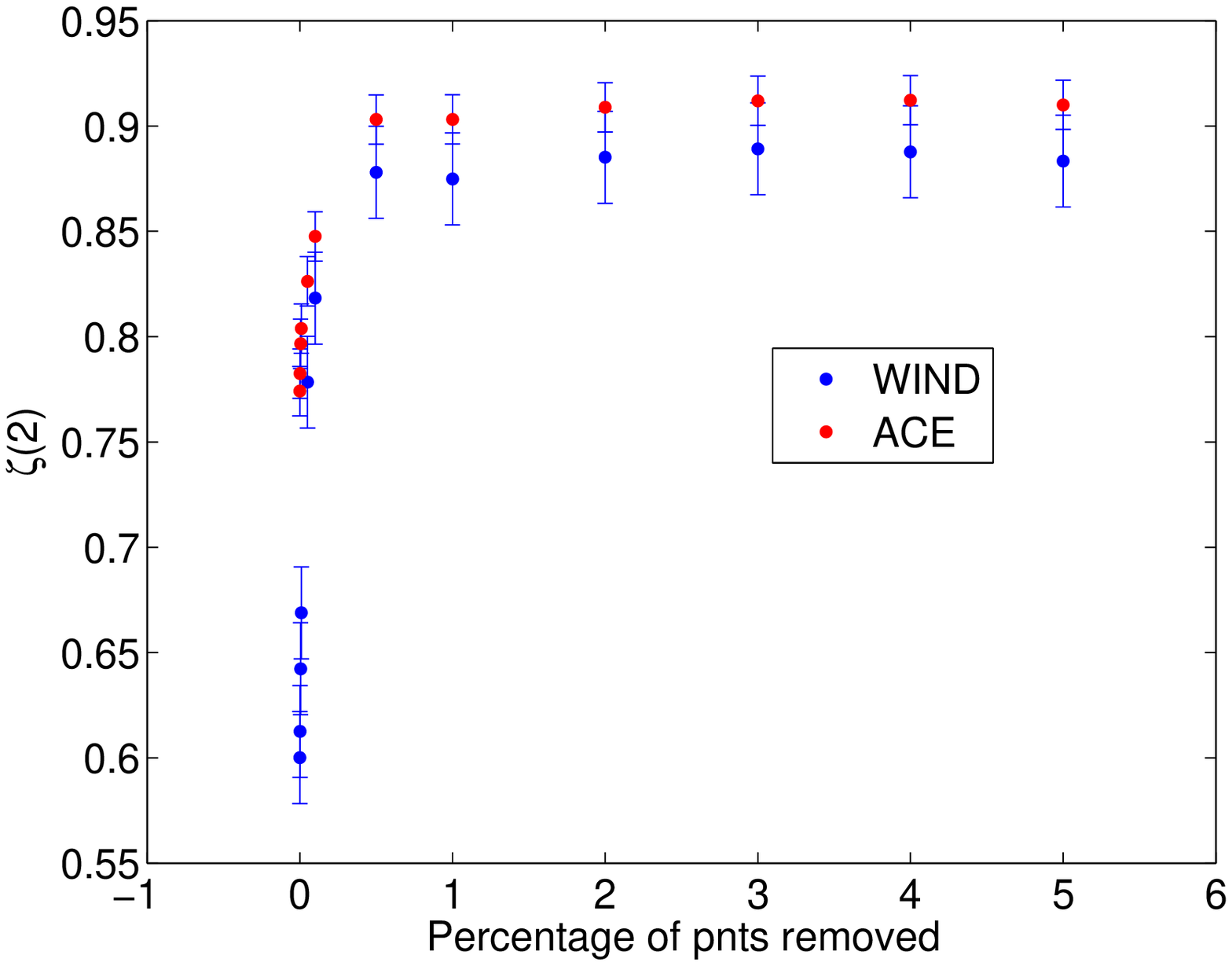}

b.)\includegraphics[width=0.8\columnwidth,keepaspectratio]{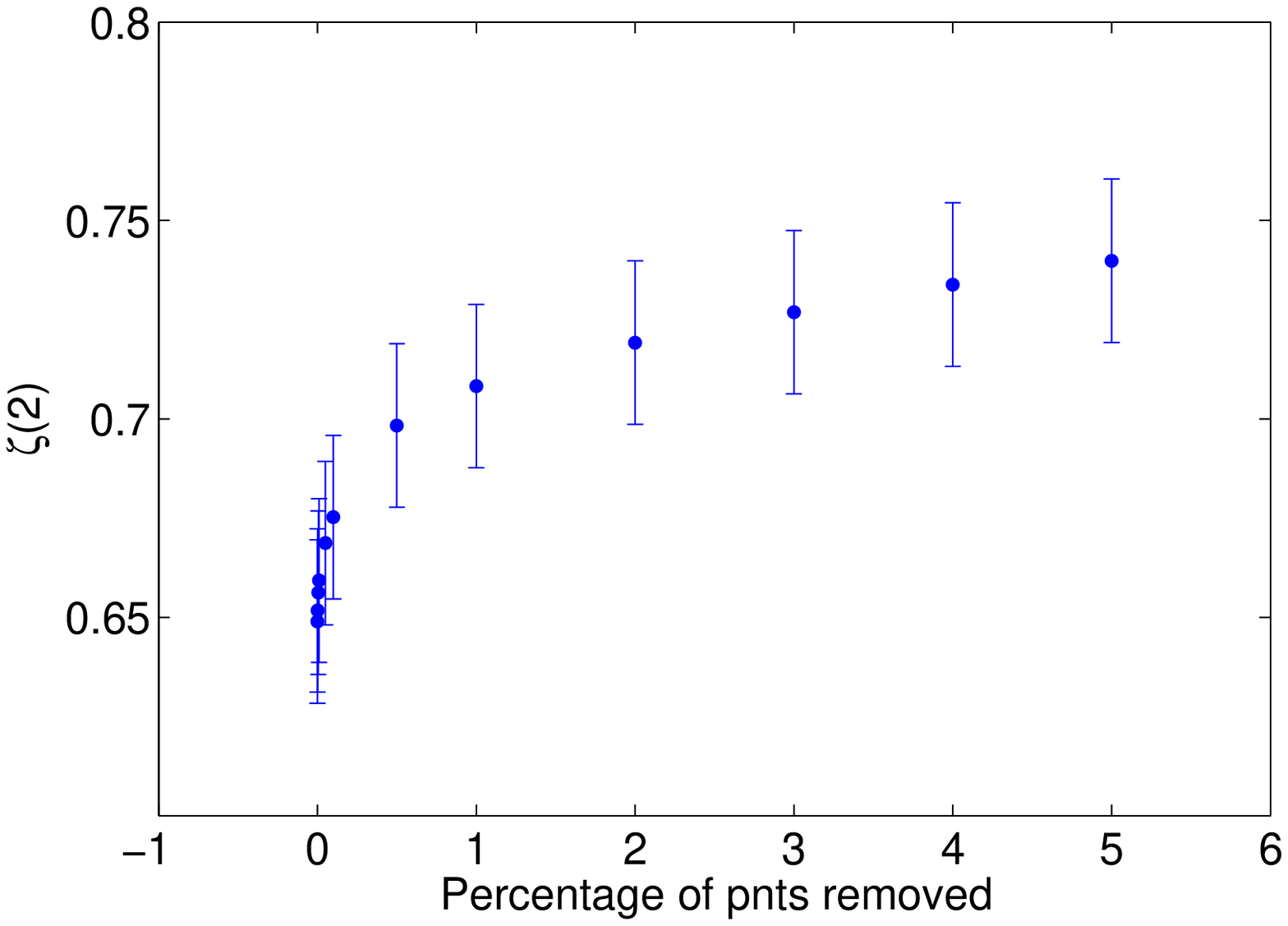}
\caption{\label{fig:zeta2VspercData}Exponent of the second moment $\zeta(2)$
Vs. the percentage of points excluded for a.) WIND
and ACE at solar maximum and b.) WIND at solar minimum.}
\end{figure}
Importantly, successively removing outlying data points does
not convert the multifractal $p$-model timeseries into a
self-affine process. In addition, a plot of $\zeta(p)$ versus
$p$ (Figure \ref{fig:zetapVsp}) is not sufficient to
distinguish self-affine from multifractal behaviour, one also
needs to examine the convergence properties of the exponents as
outlying points are successively removed, as shown in Figure
\ref{fig:zeta2Vspercsynthetic}.

We now turn to the analysis of solar wind data. In
Figures \ref{fig:zeta2VspercData}a and b we plot $\zeta(2)$ versus
the percentage of points removed in $B^{2}$ for
intervals at solar maximum and minimum respectively. The $\zeta$
values for these plots were obtained from an identified scaling
range which spanned from $\sim5.2$ minutes to $\sim2.7$ hours 
(see e.g. \cite{Hnat2003,Hnat2002} ).
Comparison of  these plots with
Figure \ref{fig:zeta2Vspercsynthetic} strongly suggests that at solar
maximum, the magnetic energy density is self-affine; we can
clearly identify a plateau with a $H=\zeta(2)/2$ value of
$H\simeq0.44\pm0.02$ for WIND and $H\simeq0.45\pm0.01$ for ACE.
At solar minimum, there is no clear plateau and the
behaviour is reminiscent of the multifractal \emph{p}-model.
We have thus differentiated the distinct scaling behaviour at
solar maximum and solar minimum. Intriguingly, it is at solar
maximum that we see self-similar behaviour; whereas at solar
minimum the timeseries resembles a multifractal, reminiscent of
intermittent turbulence. Since the corona is complex and highly
structured at solar maximum, this is highly suggestive that this
self-similar signature in $B^{2}$ is related to coronal
structure and dynamics rather than to local turbulence.

We can test this conjecture by considering observations of the solar
wind where the coronal structure is maximally ordered. We repeat
the above analysis on a two month interval of ULYSSES data during solar minimum. %
\begin{figure}
\includegraphics[width=0.9\columnwidth,keepaspectratio]{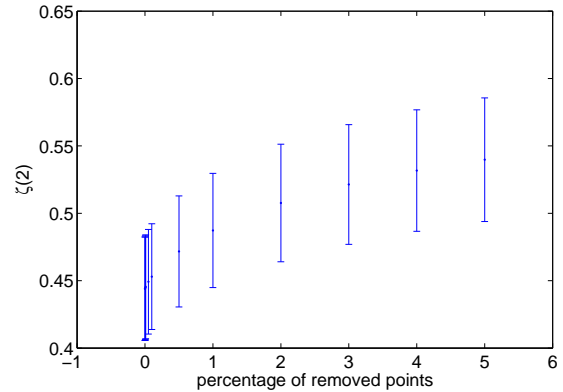}
\caption{\label{fig:zeta2VspercUlysses}Exponent of the second moment $\zeta(2)$
Vs. the percentage of points excluded for ULYSSES at solar minimum.}
\end{figure}
The resulting plot of $\zeta(2)$ versus percentage of points
excluded is shown in Figure \ref{fig:zeta2VspercUlysses}. This plot
again does not support self-affine scaling and is reminiscent of
that of the \emph{p}-model, strengthening previous results \cite{Pagel2002,Pagel2001}. 
Clearly, the behaviour of $B^2$ in the solar wind originating from a corona dominated by
ordered open field lines is not self-affine.
The appearance of fractal versus multifractal behaviour in $B^2$ is not a
strong discriminator of variability in the average solar wind speed \emph{per se}. We see 
multifractal scaling both in the ecliptic at minimum in an interval that contains periods 
of alternating high and low speed streams, and at the poles, where the average 
speed is fast and uniform. We see fractal scaling in the ecliptic at maximum, 
where the average speed also alternates between fast and slow streams.
Previous studies \cite{Pagel2002} have shown a variation with latitude and solar cycle of the 
level of multifractality of components of magnetic field. This may be related to the 
signature of the level of complexity in coronal magnetic structure which we have 
identified in $B^2$  within the inertial range of turbulence, but may also 
simply reflect a correlation with average solar wind speed.
We have also verified that $|B|$ does not show evidence 
of self-similarity for the intervals 
chosen for our study. More specifically $|B|$ exhibits multifractal behaviour. 
This confirms the earlier results of Hnat \emph{et. al.} \cite{Hnat2003}.

The corona contains many long-lived
structures which
extend far out into the solar system mediated by the
interplanetary solar wind \cite{Tu2005}. At solar maximum these structures 
show a high degree of topological complexity. One model for these
structures and their propagation is as a random walk or braiding
of magnetic field lines with a measurable diffusion coefficient
\cite{Giacalone1999,Giacalone2006,Zelenyi2004}. A diffusion
process such as this intrinsically generates self-similar scaling,
and may in a straightforward manner account for that shown here in
$B^{2}$ at solar maximum. Alternatively, the relevant
process may be that of reconnection in the complex magnetic
structure of the emerging coronal flux. Models for this include
SOC based random networks \cite{Hughes2003} which again imply
self-similar scaling. Our quantitative
determination of the Hurst exponent $H\simeq0.45$ of the self-affine 
scaling seen in the solar wind provides a strong constraint to these models.

The PDF resulting from such a self-similar process can  be captured by a solution to a
generalized Fokker-Planck equation (FPE) with power law scaling of
the transport coefficients \cite{Hnat2003,Hnat2004}. 
Intriguingly, the associated Langevin equation transforms
nonlinearly into that for a constant diffusion coefficient. The
transformation may be equivalent to introducing a
diffusion process with constant diffusion coefficient, on a space
with non-Euclidean, self-similar, fractal geometry. This may provide a
quantitative basis for models of transport of initially random
fractal fields (the coronal source) in a turbulent flow (the solar
wind). At solar minimum we see quite a different picture. Here the corona
is topologically well ordered magnetically.  Thus in this case
the fluctuations in $B^{2}$ are dominated by the evolving
turbulence of the interplanetary solar wind which is well
known to exhibit multifractal behaviour.
Intriguingly, this self-affine signature quantified here in $B^2$ 
extends over the  $\sim -5/3$ exponent inertial range seen in the solar wind. 
This is at higher frequencies than the $\sim 1/f$ behaviour previously identified as 
a coronal signature in the solar wind \cite{Matthaeus1986}. 
Although models involving reconnection
and flares and nanoflares have been proposed \cite{Velli2003}, estimates of the total 
energy contained in such structures falls significantly short of that required 
for coronal heating \cite{Aschwanden2002}. Thus the high-frequency self-similarity
reported here may suggest further processes responsible for coronal heating.

\begin{acknowledgments}
We thank N. Watkins, M. Freeman and G. Rowlands
for discussions. We acknowledge the PPARC for financial support;
R. P. Lepping and K. Oglivie for ACE and WIND data; 
and A. Balogh for the ULYSSES data.

\end{acknowledgments}

\bibliographystyle{revtex}


\end{document}